\documentclass[12pt,a4paper]{article}

\usepackage{fullpage}
\usepackage{bbm}
\usepackage{amsmath}
\usepackage{alltt, amssymb}
\usepackage{amsthm}
 \usepackage{url}
\usepackage{graphicx}

\newtheorem{theorem}{Theorem}
\newtheorem{proposition}{Proposition}[section]

\newcommand{\sfM}{{\sf M}}

\newcommand{\GFq}[1]{\mathbb{F}_{#1}}

\newcommand{\K}{\mathbf{K}}

\newcommand{\Z}{\mathbb{Z}}
\newcommand{\ZN}[1]{\Z/{#1}\Z}
\newcommand{\F}{\mathbb{F}}
\newcommand{\N}{\mathbb{N}}
\newcommand{\C}{\mathbb{C}}

\newcommand{\Es}{\tilde{E}}

\newcommand{\As}{\tilde{A}}
\newcommand{\Bs}{\tilde{B}}
\newcommand{\js}{\tilde{\jmath}}

\newcommand{\wps}{\tilde{\wp}}

\newcommand{\kerI}{F}

\usepackage{graphicx}

\title{Fast algorithms for computing isogenies \\ between elliptic
  curves \thanks{This work was supported in part by the French National Agency for Research (ANR Gecko).}}

\begin{document}

\author{
Alin Bostan, Bruno Salvy, Projet ALGO, INRIA Rocquencourt  \\
Domaine de Voluceau, 78153 Le Chesnay Cedex,  FRANCE \\
{\tt \{ Alin.Bostan, Bruno.Salvy \} @inria.fr}\\[3mm]
  Fran\c{c}ois Morain, {\'E}ric Schost, LIX, {\'E}cole polytechnique \\
  91128 Palaiseau, France\\
  {\tt \{ Francois.Morain, Eric.Schost \} @lix.polytechnique.fr} 
}

\date{\today}

\maketitle

\begin{abstract}
  We survey algorithms for computing isogenies between elliptic curves
  defined over a field of characteristic either 0 or a large prime. We
  introduce a new algorithm that computes an isogeny of degree
  $\ell$ ($\ell$ different from the characteristic) in time
  quasi-linear with respect to $\ell$. This is based in particular on
  fast algorithms for power series expansion of the Weierstrass
  $\wp$-function and related functions.
\end{abstract}



\section{Introduction}

In the Schoof-Elkies-Atkin algorithm (SEA) that computes the
cardinality of an elliptic curve over a finite field, isogenies
between elliptic curves are used in a crucial way (see for instance
\cite{BlSeSm99} and the references we give later on). Isogenies have
also been used to compute the ring of endomorphisms of a
curve~\cite{Kohel96} and isogenies of small degrees play a role
in~\cite{Galbraith99,CoHe02}.  More generally, in various contexts,
their computation becomes a basic primitive in cryptology (see
\cite{GaHeSm02,BrJo03,Smart03,DoIcKo05,JaMiVe05,Teske06,RoSt06}).

An important building block in Elkies's work is an algorithm that
computes curves that are isogenous to a given curve $E$. This block
uses modular polynomials to get the list of isogenous curves and
V\'elu's formulas to get the explicit form of the isogeny $I: E
\rightarrow \Es$, where $\Es$ is in a suitable Weierstrass form.

In this work, we concentrate on algorithms that build the degree
$\ell$ isogeny $I$ from $E$ and $\Es$ (and possibly some other
parameters, see below). For the special case $\ell=2$, formulas
exist~\cite{Silverman86}; see also \cite{CoDeMo96}. We could restrict
further to the case when $\ell$ is an odd prime, since isogenies can be written as
compositions of isogenies of prime degree, the case of prime powers
using isogeny cycles~\cite{CoMo94,CoDeMo96,FoMo02}. Besides, the odd
prime case is the most important one in SEA. However, our results
stand for arbitrary~$\ell$.  

We demand that the characteristic $p$ of the base field $\K$ be~0 or
$p \gg \ell$. This restriction is satisfied in the case of interest in
the application to the SEA algorithm, since otherwise $p$-adic methods
are much faster and easier to use~\cite{Satoh00,Kedlaya01}. Several
approaches to isogeny computation are available in small
characteristic: we refer to~\cite{Couveignes94,LeMo00} for an approach
via formal groups, \cite{Lercier96b} for the special case $p=2$, and
\cite{Couveignes96,Couveignes00,Lercier97b,JoLe06} for the general case of
$p$ small. The case of $p=\ell$ deserves a special treatment,
see~\cite{Couveignes96,LeMo00}, using Gunji's
work \cite{Gunji76} as main ingredient (see also
\cite{Lercier97b}).

Our assumption on $p$ implies that the equations of our curves can be
written in the Weierstrass form
\begin{equation}\label{curve}
y^2=x^3+Ax+B.
\end{equation}
In characteristic zero, the curve~\eqref{curve} can be parameterized by
$(x,y)=(\wp(z),\wp'(z)/2)$ in view of the classical differential equation
\begin{equation}\label{eqdiff}
\wp'(z)^2=4(\wp(z)^3+A\wp(z)+B)
\end{equation}
satisfied by the Weierstrass~$\wp$-function. This is the basis for our
computation of isogenies.  We thus prove two results, first on the
computation of the Weierstrass $\wp$-function, and then on the
computation of the isogeny itself.

Our main contribution is to exploit classical fast algorithms for
power series computations and show how they apply to the computation
of isogenies. We denote by $\sfM : \N \to \N$ a function such that
polynomials of degree less than $n$ can be multiplied in $\sfM(n)$
base field operations.  Using the fast Fourier transform~\cite{ScSt71,CaKa91},
one can take $\sfM(n) \in O(n \log n \log \log n)$; over fields
containing primitive roots of unity, one can take $\sfM(n) \in O(n\log
n)$. We make the standard super-linearity assumptions on the function
$\sfM$, see the following section.

\begin{theorem}\label{thwp}
  Let $\K$ be a field of characteristic zero. Given $A$ and $B$
  in~$\K$, the first $n$ coefficients of the Laurent expansion at the
  origin of the function~$\wp$ defined by~\eqref{eqdiff} can be
  computed in $O(\sfM(n))$ operations in $\K$.
\end{theorem}
In \S\ref{sec:wp}, we give a more precise version of this statement,
that handles the case of fields of positive, but large enough,
characteristic.

An isogeny is a regular map between two elliptic curves that is also a
group morphism.  If $E$ and $\Es$ are in Weierstrass form and
$I=(I_x,I_y)$ is an isogeny $E \to \Es$, then $I_x(P)$ depends only on
the $x$-coordinate of $P$, and there exists a constant $c \in \K$ such
that $I_y = c y I_x'$. Following Elkies~\cite{Elkies92,Elkies98}, we
consider only so-called {\it normalized} isogenies, those for which
$c=1$ (such isogenies are used for instance in SEA). In this case, we
will write $\sigma$ for the sum of the abscissas of non-zero points in
the kernel of $I$.
\begin{theorem}\label{thisog}
  Let $\K$ be a field of characteristic~$p$ and let $E$ and $\Es$ be
  two curves in Weierstrass form, such that there exists a normalized
  isogeny $I: E \rightarrow \Es$ of degree~$\ell$. Then, one can
  compute the isogeny $I$
\begin{enumerate}
\item in $O(\sfM(\ell))$ operations in~$\K$, if $p=0$ or $p>2\ell-1$, if
  $\sigma$ is known;
\item in $O(\sfM(\ell) \log \ell)$ operations in~$\K$, if $p=0$ or
  $p>8\ell-5$, without prior knowledge of~$\sigma$.
\end{enumerate}
\end{theorem}
Taking ${\sf M}(n) \in O(n \log n \log \log n)$ shows that the
complexity results in Theorems~\ref{thwp} and~\ref{thisog} are nearly
optimal, up to polylogarithmic factors. 
Notice that the algorithms using modular equations to detect isogenies
yield the value of $\sigma$ as a by-product.  However, in a
cryptographic context, this may not be the case anymore; this is why
we distinguish the two cases in Theorem~\ref{thisog}.

This article is organized as follows. In \S\ref{section:Newton}, we
recall known results on the fast computation of truncated power
series, using notably Newton's iteration. In \S\ref{sec:wp}, we show
how these algorithms apply to the computation of
the~$\wp$-function. Then in \S\ref{section:new}, we recall the
definition of isogenies and the properties we need, and give our
quasi-linear algorithms; examples are given in~\S\ref{sct:examples}.
In the next section, we survey previous algorithms for the computation
of isogenies. Their complexity has not been discussed before; we
analyze them when combined with fast power series expansions so that a
comparison can be made. Finally, in \S\ref{section:implementation}, we
report on our implementation.


\section{A review of fast algorithms for power series}\label{section:Newton}

The algorithms presented in this section are well-known; they reduce
several problems for power series (reciprocal, exponentiation,~\dots)
to polynomial multiplication.

Our main tool to devise fast algorithms is Newton's iteration; it
underlies the $O(\sfM(\ell))$ result reported in Theorem~\ref{thwp},
and in the (practically important) point~(1) of Theorem~\ref{thisog}.
Hence, this question receives most of our attention below, with
detailed pseudo-code. We will be more sketchy on some other
algorithms, such as rational function reconstruction, referring to the
relevant literature.

We suppose that the \emph{multiplication time} function $\sfM$ is
super-linear, {\it i.e.}, it satisfies the following inequality (see, e.g.,
\cite[Chapter 8]{GaGe99}): 
\begin{equation}\label{eq:M}
\frac{\sfM(n)}n \leq \frac{\sfM(n')}{n'}
\qquad \text{~if~} n \leq n'. 
\end{equation}
In particular, Equation~\eqref{eq:M} implies the inequality
\begin{equation*}
\sfM(1) + \sfM(2) +\sfM(4)+ \cdots + \sfM(2^i) \leq 2 \sfM(2^i),
\end{equation*}
which is the key to show that all algorithms based on Newton's
iteration have complexity in $O(\sfM(n))$. Cantor and
Kaltofen~\cite{CaKa91} have shown that one can take $\sfM(n)$ in $O(n
\log n \log \log n)$; as a byproduct, most questions addressed below
admit similar quasi-linear estimates.


\subsection{Reciprocal}\label{ssec:inv}
Let $f=\sum_{i \geq 0} f_i z^i$ be in $\K[[z]]$, with $f_0\neq 0$, and
let $g = 1/f =\sum_{i \geq 0} g_i z^i$ in $\K[[z]]$. The coefficients
$g_i$ can be  computed  iteratively by the formula
$$g_0 = \frac{1}{f_0} \quad \text{and} \quad g_i = -\frac{1}{f_0}
\sum_{j=1}^{i} f_j g_{i-j} \quad \text{~for~} \quad i \geq 1.$$ For a
general $f$, the cost of computing $1/f \bmod z^n$ with this method is
in $O(n^2)$; observe nevertheless that if $f$ is a polynomial of
degree $d$, the cost reduces to $O(nd)$.

To speed up the computation in the general case, we use Newton's
iteration. For reciprocal computation, it amounts to computing a
sequence of truncated power series $h_i$ as follows:
$$h_0 = \frac{1}{f_0} \quad \text{and} \quad h_{i+1} = h_i(2-f h_i)
\bmod z^{2^{i+1}} \text{~for~} i \geq 0 .$$ Then, $h_i = 1/f 
\bmod z^{2^i}.$ As a consequence, $1/f \bmod z^n$ can be computed in
$O(\sfM(n))$ operations. This result is due to Cook for an analogous
problem of integer inversion~\cite{Cook66}, and to
Sieveking~\cite{Sieveking72} and Kung~\cite{Kung74} in the power
series case.


\subsection{Exponentiation}\label{ssec:exp} 
Let $f$ be in $\K[[z]]$, with $f(0)=0$.  Given $n$ in $\N$, such that
$2,\dots,n-1$ are units in $\K$, the truncated exponential $\exp_n(f)$
is defined as
$$\exp_n(f) = \sum_{i=0}^{n-1} \frac1{i!}f^i \bmod z^{n}.$$
Conversely, if $g$ is in $1+z\K[[z]]$, its truncated logarithm is
defined as
$$\log_{n}(g) = -\sum_{i=1}^{n-1} \frac1{i}(1-g)^i \bmod z^{n}.$$ The
truncated logarithm is obtained by computing the Taylor expansion of
${g'}/g$ modulo $z^{n-1}$ using the algorithm of the previous
subsection, and taking its antiderivative; hence, it can be computed
in $O(\sfM(n))$ operations.

Building on this, Brent~\cite{Brent75} introduced the Newton iteration
$$g_0 = 1, \quad g_{i+1} = g_i(1+f-\log_{2^{i+1}}(g_i)) \bmod
z^{2^{i+1}}$$ to compute the sequence $g_i = \exp_{2^i}(f)$. As a
consequence, $\exp_n(f)$ can be computed in $O(\sfM(n))$ operations
as well, whereas the naive algorithm has cost $O(n^2)$.

As an application, Sch\"onhage~\cite{Schonhage82} gave a fast
algorithm to recover a polynomial $f$ of degree $n$ from its first $n$
power sums $p_1,\dots,p_n$. Sch\"onhage's algorithm is based on the
fact that the logarithmic derivative of $f$ at infinity is the
generating series of its power sums, that is,
$$z^n f \left ( \frac 1z \right ) = \exp_{n+1}\!\left (-{\sum_{i=1}^n
    \frac{p_i}i z^{i}}\right).$$ Hence, given $p_1,\dots,p_n$, the
coefficients of $f$ can be recovered in time $O(\sfM(n))$. This
algorithm requires that $2,\dots,n$ be units in $\K$.


\subsection{First-order linear differential equations}\label{ssec:linear}
Let $a,b,c$ be in $\K[[z]]$, with $a(0) \neq 0$, and let $\alpha$ be
in $\K$. We want to compute the first~$n$ terms of $f \in \K[[z]]$
such that
$$af'+bf =c \quad \text{and} \quad f(0)=\alpha$$ Let $B= b/a \bmod
z^{n-1}$ and $C= c/a \bmod z^{n-1}$. Then, defining $J=\exp_n
(\int{B})$, $f$ satisfies the relation
$$f = \frac 1J\left (\alpha + \int CJ\right ) \bmod z^n.$$ Using the
previous reciprocal and exponentiation algorithms, $f \bmod z^n$ can
thus be computed in time $O(\sfM(n))$. This algorithm is due to Brent
and Kung~\cite{BrKu78}; it requires that $2,\dots,n-1$ be units in $\K$.


\subsection{First-order nonlinear differential equations}\label{ssec:R}
We only treat this question in a special case, following again Brent
and Kung's article~\cite[Theorem~5.1]{BrKu78}. Let $G$ be in
$\K[[z]][t]$, let $\alpha, \beta$ be in $\K$, and let $f\in \K[[z]]$
be a solution of the equation
$$f'^2 = G(z,f),\quad f(0) = \alpha, \quad f'(0)=\beta,$$ 
with furthermore $\beta^2=G(0,\alpha)\neq 0$. Supposing that, for $s
\geq 2$, the initial segment $f_1=f \bmod z^s$ is known, we show how
to deduce $f \bmod z^{2s-1}$.  Write $f = f_1+f_2 \bmod z^{2s-1}$,
where $z^s$ divides~$f_2$. One checks that $f_2$ is a solution of the
linearized equation
\begin{equation}\label{linearized}
  2 f_1' f_2' - G_t(z,f_1)f_2 = G(z,f_1) - f_1'^2 \mod z^{2s-2},
\end{equation}
with the initial condition $f_2(0)=0$, where~$G_t$ denotes the
derivative of~$G$ with respect to $t$. The condition $f'(0) \neq 0$
implies that $f_1'$ is a unit in $\K[[z]]$; then, the cost of computing
$f_2 \bmod z^{2s-1}$ is in $O(\sfM(s))$ (remark that we do not take
the degree of $G$ into account). Finally, the computation
of~$f$ at precision~$n$ is as follows:
\begin{enumerate}
\item Let $f=\alpha+\beta z\bmod z^2$ and $s=2$;
\item {\bf while} $s < n$ {\bf do}
\begin{enumerate}
\item Compute $f \bmod z^{2s-1}$ from $f \bmod z^s$;
\item Let $s = 2s-1$.
\end{enumerate}
\end{enumerate}
Due to the super-linearity of $\sfM$, $f \bmod z^n$ can thus be
computed using $O(\sfM(n))$ operations. Again, we have to assume that
$2,\dots,n-1$ are units in $\K$.


\subsection{Other algorithms.}\label{ssec:other}
We conclude this section by pointing out other algorithms that are
used below.

\medskip
\paragraph{\sc \small Power series composition.} Over a general field
$\K$, there is no known algorithm of quasi-linear complexity for
computing $f(g) \bmod z^n$, for $f,g$ in $\K[[z]]$. The best results
known today are due to Brent and Kung~\cite{BrKu78}. Two algorithms
are proposed in that article, of respective complexities
$O(\sfM(n)\sqrt{n} + n^{\frac{\omega+1}2})$ and $O(\sfM(n) \sqrt{n
  \log n})$, where $2\leq \omega <3$ is the exponent of matrix
multiplication (see, e.g., \cite[Chapter~12]{GaGe99}). Over
fields of positive characteristic~$p$, Bernstein's algorithm for
composition~\cite{Bernstein98} has complexity $O(\sfM(n))$, but the
$O(\,)$ estimate hides a linear dependence in~$p$, making it
inefficient in our setting ($p \gg n$).

\medskip
\paragraph{\sc \small Rational function reconstruction.} Our last
subroutine consists in reconstructing a rational function from its
Taylor expansion at the origin. Suppose that $f$ is in $\K(z)$ with
numerator and denominator of degree bounded respectively by $n$ and
$n'$, and with denominator non-vanishing at the origin; then, knowing
the first $n+n'+1$ terms of the expansion of $f$ at the origin, the
rational function $f$ can be reconstructed in $O(\sfM(n+n')\log
(n+n'))$ operations, see~\cite{BrGuYu80}.


\section{Computing the Weierstrass $\wp$-function}\label{sec:wp}


\subsection{The Weierstrass $\wp$-function}
We now study the complexity of computing the Laurent series expansion
of the Weierstrass $\wp$-function at the origin, thus proving
Theorem~\ref{thwp}. We suppose for a start that the base field $\K$
equals $\C$; the positive characteristic case is discussed below.  Let
thus $A,B$ be in $\K=\C$. The Weierstrass function $\wp$ associated to
$A$ and $B$ is a solution of the non-linear differential
equation~\eqref{eqdiff}; its Laurent expansion at the origin has the
form
\begin{equation}\label{wpdev}
  \wp(z) = \frac{1}{z^2} + \sum_{i \geq 1} c_i z^{2i}.
\end{equation}
The goal of this section is to study the complexity of computing the
first terms $c_1,\dots,c_n$.  We first present a ``classical''
algorithm, and then show how to apply the fast algorithms for power
series of the previous section.


\subsection{Quadratic algorithm} \label{ssec:c} First, we recall the
direct algorithm. Substituting the expansion~\eqref{wpdev} into
Equation~\eqref{eqdiff} and identifying coefficients of $z^{-2}$ and
$z^0$ gives
$$c_1 = -\frac A5 \quad \text{and} \quad c_2 = -\frac B7.$$ Next,
differentiating Equation~\eqref{eqdiff} yields the second order
equation
\begin{equation}\label{eqdifff} 
\wp'' = 6 \wp^2 + 2 A.
\end{equation}
This equation implies that for $k\geq 3$, $c_k$ is given by
\begin{equation}\label{eqc} 
c_k = \frac{3}{(k-2) (2k+3)} \sum_{i=1}^{k-2} c_i c_{k-1-i}.
\end{equation}
Hence, the coefficients $c_1,\dots,c_n$ can be computed using $O(n^2)$
operations in $\K$.

If the characteristic $p$ of $\K$ is positive, the definition of $\wp$
as a Laurent series fails, due to divisions by zero. However, assuming
$p > 2n+3$, it is still possible to define the coefficients
$c_1,\dots,c_n$ through the previous recurrence relation. Then, again,
$c_1,\dots,c_n$ can be computed using $O(n^2)$ operations in $\K$.

\subsection{Fast algorithm} \label{fastwp}  
We first introduce new quantities, that are used again in the next
section.  Define
$$Q(z) = \frac 1{\wp(z)} \in z^2+z^6\K[[z^2]] \quad \text{and} \quad
R(z) = \sqrt{Q(z)} \in z+z^5\K[[z^2]].$$ The differential equation
satisfied by $R$ is
\begin{equation}\label{eqdiffR}
  {R'}(z)^2 = B \, R(z)^6 + A\, R(z)^4 + 1,
\end{equation}
from which we can deduce the first terms of $R$:
$$R(z) = z + \frac{A}{10} z^5 + \frac{B}{14} z^7 + O(z^8) = z \left( 1 + \frac{A}{10} z^4 + \frac{B}{14} z^6 + 
  O(z^7)\right).$$ Squaring $R$ yields
$$Q(z) = z^2 + \frac{A}{5} z^6 + \frac{B}{7} z^8 + O(z^{9})
= z^2 \left (1 + \frac{A}{5} z^4 + \frac{B}{7} z^6 + O(z^7)\right) .$$
Taking the reciprocal of the right-hand series finally yields
$$\wp(z) = \frac 1{z^2}\left (1 - \frac{A}{5} z^4 - \frac{B}{7} z^6 + O(z^7) \right) = \frac 1{z^2} 
- \frac{A}{5} z^2 - \frac{B}{7} z^4 + O(z^5),$$ as requested. Thus,
our fast algorithm to compute the coefficients $c_1,\dots,c_n$ is as
follows:
\begin{enumerate}
\item Compute~$R(z)\bmod z^{2n+4}$ using the algorithm of
  \S\ref{ssec:R} with $G=Bt^6+At^4+1$;
\item Compute~$Q(z)=R(z)^2\bmod z^{2n+5}$;
\item Compute~$\wp(z)=1/Q(z)\bmod z^{2n+1}$.
\end{enumerate}
In the first step, we remark that our assumption $R'(0) \neq 0$ is
indeed satisfied, hence $R(z) \bmod z^{2n+4}$ can be computed in
$O(\sfM(n))$ operations, assuming $2,\dots,2n+3$ are units in $\K$.
Using the algorithm of \S\ref{ssec:inv}, the squaring and reciprocal
necessary to recover $\wp(z) \bmod z^{2n+1}$ admit the same complexity
bound. This proves Theorem~\ref{thwp}.


\section{Fast computation of isogenies}\label{section:new}

In this section, we recall the basic properties of isogenies and an
algorithm due to Elkies~\cite{Elkies98} that computes an isogeny of
degree~$\ell$ in quadratic complexity $O(\ell^2)$. Then, we design two
fast variants of Elkies' algorithm, by exploiting the differential
equations satisfied by some functions related to the Weierstrass
function, proving Theorem~\ref{thisog}.


\subsection{Isogenies}
The following properties are classical; all the ones not proved here
can be found for instance in~\cite{Silverman86,Silverman94}.
Let $E$ and $\Es$ be two elliptic curves defined over $\K$. An isogeny
between $E$ and $\Es$ is a regular map $I: E \to \Es$ that is also a
group morphism. Hence, we have $\Es \simeq E/\kerI$, where $\kerI$ is
the kernel of $I$; here, our isogenies are all non-zero.

The most elementary example of an isogeny is the ``multiplication by
$m$'' map which sends $P \in E$ to $[m]P$, where, as usual, the group
law on $E$ is written additively. If $E$ is given through a
Weierstrass model, the group law yields the following formulas for
$[m]P$ in terms of the \emph{Weber polynomials} $\psi_m(x, y)$
\cite[p.~105]{Silverman86}:
\begin{equation}\label{eq:divpol}
[m](x, y) = \left(\frac{\phi_m(x, y)}{\psi_m(x, y)^2}, \frac{\omega_m(x,
y)}{\psi_m(x, y)^3}\right).
\end{equation}
Using the Weierstrass equation of $E$, the polynomial $\psi_m(x, y)$
rewrites in terms of the so-called \emph{division polynomial}
$f_m(x)$, which is univariate of degree $\Theta(m^2)$: $\psi_m(x, y) =
f_m(x)$ if $m$ is odd, $\psi_m(x, y) = 2 y f_m(x)$ otherwise.

Given an isogeny $I: E \rightarrow \Es$, there exist a unique isogeny
(the {\em dual isogeny}) $\hat{I}: \Es \rightarrow E$ and a unique
integer $\ell$ such that $\hat{I} \circ I = [\ell]$; the integer
$\ell$ is called the {\em degree} of $I$; if $I$ is separable, it
equals the cardinality of its kernel. For instance, the degree of the
isogeny~$[m]$ is $m^2$ and this is reflected by the degree of the
division polynomials.

Let $E$ and $\Es$ be two isogeneous elliptic curves in Weierstrass
form, defined over $\K$. Then the isogeny $I$ between $E$ and $\Es$
can be written as
\begin{equation}\label{Dewaghe:0}
  I(x, y) = \left(I_x(x), c y  I'_x(x)\right),
\end{equation}
for some $c$ in $\K$. We say that $I$ is {\em normalized\/} if the
constant $c$ equals $1$; in this case, $I$ is separable.  We use an
explicit form for such isogenies, extending results of
Kohel~\cite[\S2.4]{Kohel96} and Dewaghe~\cite{Dewaghe99} to the case
of arbitrary degree $\ell$.

\begin{proposition}\label{prop:dewaghe}
Let $I: E \to \Es$ be a normalized isogeny of degree $\ell$ and let
$\kerI$ be its kernel. Then $I$ can be written as
\begin{equation}\label{Dewaghe}
  I(x, y) = \left(\frac{N(x)}{D(x)}, y
    \left(\frac{N(x)}{D(x)}\right)'\right),
\end{equation}
where $D$ is the polynomial
\begin{equation}\label{eq:D}
D(x) = \prod_{Q\in\kerI^*} (x - x_Q) = x^{\ell-1} - \sigma
x^{\ell-2} + \sigma_2 x^{\ell-3} - \sigma_3 x^{\ell-4}+\cdots
\end{equation}
and $N(x)$ is related to $D(x)$ through the formula
\begin{equation}\label{eqND}
  \frac{N(x)}{D(x)}=\ell x-\sigma - (3 x^2+A)
    \frac{D'(x)}{D(x)}-2(x^3+Ax+B) \left ( \frac{D'(x)}{D(x)} \right )'.
\end{equation}
\end{proposition}
\begin{proof}[{\sc Proof.}]
Note first that given a subgroup $\kerI$ of $E(\overline \K)$, there
can exist only one pair $(\Es,I)$ where $\Es$ is in Weierstrass form
and $I$ is a normalized isogeny $E \to \Es$ having $\kerI$ as kernel.

In~\cite{Velu71}, V\'elu constructs the curve $\Es$ and the normalized
isogeny $I$, starting from the coordinates of the points in its kernel
$\kerI$. A point $P$ of coordinates $(x_P,y_P)$ is sent by the
isogeny~$I$ to a point of coordinates
$$x_{I(P)}=x_P+\sum_{Q\in \kerI^*}(x_{P+Q}-x_Q) \quad\text{and}\quad
y_{I(P)}=y_P+\sum_{Q\in \kerI^*}(y_{P+Q}-y_Q).$$ From there, V\'elu uses the group law to get explicit expressions of the coordinates. More precisely,
write $\kerI_2$ for
the set of points in $\kerI$ that are of order $2$. Then $\kerI$ can
be written as $$F=\{O_E\} \cup \kerI_2 \cup \kerI_{\rm odd} \cup
(-\kerI_{\rm odd}),$$ where $\kerI_{\rm odd} \cap (-\kerI_{\rm odd}) =
\emptyset$ and $-\kerI_{\rm odd}$ denotes the set of opposite points
of $\kerI_{\rm odd}$, so that $D(x)$ rewrites as
$$D(x) = \prod_{Q\in\kerI_2} (x - x_Q) \prod_{Q\in\kerI_{\rm odd}} (x
- x_Q)^2.$$ Finally, let $\kerI^{+} = \kerI_2 \cup \kerI_{\rm
odd}$. Then V\'elu gave the following explicit form for $I(x,y)=(I_x(x), y
I_x'(x))$:
\begin{eqnarray*}
  I_x(x) &=&
    x + {\small\sum_{Q\in \kerI^{+}}} \left( \frac{t_Q}{x-x_Q} + 4
      \frac{x_Q^3+Ax_Q + B}{(x-x_Q)^2} \right), 
\end{eqnarray*}
where $t_Q = 3 x_Q^2+A$ if $Q\in\kerI_2$ and $t_Q=2 (3 x_Q^2+A)$
otherwise. Observing that for $Q \in \kerI_2$, $x_Q^3+Ax_Q + B$ equals $0$, one
sees that $I_x$ admits $D$ for denominator, as claimed in
Equation~\eqref{Dewaghe}.

Next, we split the sum over $\kerI^+$ into that for $Q\in \kerI_2$ and
that for $Q \in \kerI_{\rm odd}$. The former rewrites as $$\sum_{Q\in
\kerI_2} \left( \frac{3 x_Q^2+A}{x-x_Q} + 2 \frac{x_Q^3+Ax_Q +
B}{(x-x_Q)^2} \right),$$ since these points satisfy $x_Q^3+A x_Q+B=0$;
the sum for $Q \in \kerI_{\rm odd}$ rewrites as
$$ \frac{1}{2} \sum_{Q \in \kerI_{\rm odd} \cup\,
-\kerI_{\rm odd}}
\left( \frac{t_Q}{x-x_Q} + 4
      \frac{x_Q^3+Ax_Q + B}{(x-x_Q)^2} \right).$$
Therefore, we obtain
$$I_x(x) = x + \sum_{Q \in \kerI^*} \left(\frac{3x_Q^2 + A}{x-x_Q} + 2
      \frac{x_Q^3+Ax_Q + B}{(x-x_Q)^2} \right),$$
which can be rewritten as
\begin{eqnarray*}
I_x(x) &=& x + \sum_{Q \in F^*}\left( x-x_Q -
    \frac{3 x^2+A}{x-x_Q}+2 \frac{x^3+A x +B}{(x-x_Q)^2}  \right).
\end{eqnarray*}
This yields Equation~\eqref{eqND}.
\end{proof}

\smallskip

Though this is not required in what follows, let us mention how
V\'elu's formul\ae{} enable one to construct the curve $\Es$. 
Let $\sigma,\sigma_2,\sigma_3$ be as in Equation~\eqref{eq:D}
and 
$$t = A (\ell-1) + 3 (\sigma^2 - 2\sigma_2),$$
$$w = 3 A \sigma + 2 B (\ell-1) + 5 (\sigma^3
-3\sigma\sigma_2+3\sigma_3).$$ Then the isogenous curve $\Es$ has the
Weiestrass equation $Y^2 = X^3+\As X +\Bs$, where $\As = A - 5 t$ and
$\Bs = B - 7 w$.


The constant $\sigma$ introduced in the previous proposition is the
sum of the abscissas of the points in the kernel $\kerI$ of $I$. In
the important case where $\ell$ is odd, the non-zero points in $\kerI$
come into pairs $\{(x_Q,y_Q), (x_Q,-y_Q)\}$, so that we will write
$$D(x) = g(x)^2 \quad\text{with}\quad g(x) = x^{(\ell-1)/2} - q_1
x^{(\ell-3)/2} + \cdots,$$ and $\sigma = 2 q_1$. Then, we can replace
${D'(x)}/{D(x)}$ by $2 {g'(x)}/{g(x)}$ in 
Proposition~\ref{prop:dewaghe}.


\medskip
\subsection{Elkies' quadratic algorithm}~\label{elkies98} 
{}From now on, we are given the two curves $E$ and $\Es$ through their
Weierstrass equations, admitting a normalized isogeny $I:E\to\Es$ of
degree $\ell$. We will write 
\begin{equation*}
E:\ y^2=x^3+Ax+B
\quad\text{and}\quad
\Es:\ y^2=x^3+\As x+ \Bs.
\end{equation*}
{}From this input, and possibly
that of $ \sigma$, we want to determine the isogeny $I$, which we
write as in Equation~\eqref{Dewaghe}
\begin{equation*}
  I(x, y) = \left(\frac{N(x)}{D(x)}, y
    \left(\frac{N(x)}{D(x)}\right)'\right).
\end{equation*}
We first describe an algorithm due to Elkies~\cite{Elkies98}, that we
call {\sf Elkies1998}, whose complexity is quadratic in the
degree~$\ell$. In the next subsection, we give two fast variants of
algorithm {\sf Elkies1998}, called {\sf fastElkies} and {\sf
fastElkies$'$}, of respective complexities $O(\sf M(\ell))$ and $O(\sf
M(\ell) \log \ell)$.

The algorithm {\sf Elkies1998} was introduced for the prime degree
case in~\cite{Elkies98}, but it works for any $\ell$ large enough. The
first part of the algorithm aims at computing the expansion of
$N(x)/D(x)$ at infinity; the second part amounts to recovering the
power sums of the roots of~$D(x)$ from this expansion.

To present these ideas, our starting remark is that the rational
function $N(x)/D(x)$ satisfies the non-linear differential equation
\begin{equation}\label{eq:ND}
  (x^3+Ax+B) {\left ( \frac {N(x)}{D(x)} \right )'}^{\,2}=
  \left ( \frac {N(x)}{D(x)} \right )^3+
  \tilde{A} \left (\frac {N(x)}{D(x)}\right )+\tilde{B}.
\end{equation}
This follows from Proposition~\ref{prop:dewaghe} and the fact that $I$
maps $E$ onto $\Es$. Differentiating Equation~\eqref{eq:ND} leads to
the following second-order equation:
\begin{equation}\label{eq:diff}
  (3 x^2 + A) \left (\frac {N(x)}{D(x)}  \right )' 
  + 2 (x^3+A x +B) \left ( \frac {N(x)}{D(x)} \right )''
  = 
  3 \left (\frac {N(x)}{D(x)} \right )^2 + \As .\end{equation} 
Writing the expansion of the rational function $N(x)/D(x)$ at infinity 
$$\frac{N(x)}{D(x)} = x + \sum_{i\geq 1} \frac{h_i}{x^i}$$ and
identifying coefficients of $x^{-i}$ from both sides of
Equation~\eqref{eq:diff} yields the recurrence 
\begin{equation}\label{eq:elkies98}
h_{k} = \frac{3}{(k-2)(2k+3)} \sum_{i=1}^{k-2} h_i h_{k-1-i} - \frac{2k-3}{2k+3}
A h_{k-2} -\frac{2(k-3)}{2k+3} B  h_{k-3}, \quad \text{for all} \; k\geq 3, 
\end{equation}
with  initial conditions
$$ h_1 = \frac{A-\As}5 \quad \text{and}\quad h_2 = \frac{B-\Bs}7.$$
The recurrence~\eqref{eq:elkies98} is the basis of algorithm {\sf
  Elkies1998}; using it, one can compute $h_3,\dots,h_{\ell-2}$
using $O(\ell^2)$ operations in~$\K$.

Elkies' algorithm {\sf Elkies1998} assumes that~$\sigma$ is given.
Extracting coefficients in Equation~\eqref{eqND} then yields
\begin{equation}\label{fnpn}
h_i = (2i+1) p_{i+1} + (2i-1) A p_{i-1} + (2i-2) B p_{i-2}, \quad \text{for all} \;   i \geq 1.
\end{equation}
Since $h_1,\dots,h_{\ell-2}$ are known,~$p_2,\dots,p_{\ell-1}$ can be
deduced from the previous recurrence using $O(\ell)$ operations. The
polynomial $D(x)$ is then recovered, either by a quadratic algorithm
or the faster algorithm of~\S\ref{ssec:exp}, and $N(x)$ is deduced
using formula~\eqref{eqND}, in $O(\sfM(\ell))$ operations.

This algorithm requires that $2,\dots,2\ell-1$ be units in $\K$.  Its
complexity is in $O(\ell^2)$, the bottleneck being the computation of
the coefficients $h_1,\dots,h_{\ell-2}$. Observe the parallel with the
computations presented in the previous section, where differentiating
Weierstrass' equation yields the recurrence~\eqref{eqc}, which appears
as a particular case of the recurrence~\eqref{eq:elkies98} (the former
is obtained by taking $A=B=0$ in the latter).

\medskip
\subsection{Fast algorithms}~\label{fastElkies} We improve on the
computation of the coefficients $h_i$ in algorithm {\sf Elkies1998},
the remaining part being unchanged.  Unfortunately, we cannot directly
apply the algorithm of \S\ref{ssec:R} to compute the expansion of
$N(x)/D(x)$~at infinity using the differential equation~\eqref{eq:ND},
since the equation obtained by the change of variables $x \mapsto 1/x$
is singular at the origin. To avoid this technical complication, we
rather consider the power series
$$S(x) = x + \frac{\tilde{A} - A}{10} x^5 +\frac{\tilde{B} - B}{14}
x^7 +O(x^9) \in x + x^3\K[[x^2]]$$ such that 
$$\frac{N(x)}{D(x)} = \frac{1}{S \left(\frac{1}{\sqrt{x}} \right)
  ^2};$$ remark that $S$ satisfies the relation $\tilde{R} = S \circ
  R$, with the notation $R(z) =1 / \sqrt{\wp(z)}$ and $ \tilde{R}(z)
  =1 / \sqrt{\wps(z)}$ introduced in~\S\ref{fastwp}.

Applying the chain rule gives the following first order differential
equation satisfied by~$S(x)$:
$$(B x^6 + A x^4 + 1)\, {S{\,}'(x)^{2}} = 1 + \tilde{A}\, S(x)^4 + \tilde{B}\,
S(x)^6.$$ 
Using this differential equation, we propose two algorithms to compute
$N(x)/D(x)$, depending on whether the coefficient $\sigma$ is known or not. In
the algorithms, we write
$$S(x) = xT(x^2) \quad \text{and} \quad U(x) = \frac{1}{T(x)^2} \in 1 + x^2\K[[x]]
\quad \text{so that} \quad \frac{N(x)}{D(x)} = x\, U\left( \frac{1}{x}
\right ).$$ The first algorithm, called {\sf fastElkies}, assumes that
$\sigma$ is known and goes as follows. 
\begin{enumerate}
\item Compute $C(x) = (B x^6 + A x^4 + 1)^{-1} \bmod x^{2\ell-1} \in \K[[x]]$;
\item Compute~$S(x) \bmod x^{2\ell}$ using the algorithm of
  \S\ref{ssec:R} with $G(x,t) = C(x) (1 + \tilde{A} t^4 + \tilde{B}
  t^6)$, and deduce $T(x) \bmod x^{\ell}$;
\item Compute~$U(x)=1/T(x)^2\bmod x^{\ell}$ using the algorithm in~\S\ref{ssec:inv};
\item Compute the coefficients $h_1,\ldots, h_{\ell-2}$ of~$N(x)/D(x)$, using
  $N(x)/D(x) = x U(1/x)$;
\item Compute the power sums $p_2,\ldots,p_{\ell-1}$ of $D(x)$, using the
  linear recurrence~\eqref{fnpn};
\item Recover $D(x)$ from its power sums, as described in~\S\ref{ssec:exp};
\item Deduce $N(x)$ using Equation~\eqref{eqND}.
\end{enumerate}
Steps (1) and (5) have cost $O(\ell)$. Steps (2), (3), (6) and (7) can
be performed in~$O(\sfM(\ell))$ operations, and Step (4) requires no
operation. This proves the first part of Theorem~\ref{thisog}.

\medskip For our second algorithm, that we call {\sf fastElkies$'$},
we do not assume prior knowledge of $\sigma$. Its steps (1')--(3')
are just a slight variation of Steps (1)--(3), of the same
complexity~$O(\sfM(\ell))$, up to constant factors.
\begin{enumerate}
\item[(1')] Compute $C(x) = (B x^6 + A x^4 + 1)^{-1} \bmod x^{8\ell-5}
  \in \K[[x]]$;
\item[(2')] Compute~$S(x) \bmod x^{8\ell-4}$ using the algorithm of
  \S\ref{ssec:R} with $G(x,t) = C(x) (1 + \tilde{A} t^4 + \tilde{B}
  t^6)$, and deduce $T(x) \bmod x^{4\ell-2}$;
\item[(3')] Compute~$U(x)=1/T(x)^2\bmod x^{4\ell-2}$, using the algorithm
  in~\S\ref{ssec:inv};
\item[(4')] Reconstruct the rational function $U(x)$;
\item[(5')] Return $N(x)/D(x) = x U(1/x)$.
\end{enumerate}
Using fast rational reconstruction, Step (4') can be performed in
$O({\sf M}(\ell) \log \ell)$ operations in~$\K$. Finally, it is easy
to check that our algorithm {\sf fastElkies} requires that
$2,\ldots,2\ell-1$ be units in~$\K$, while algorithm {\sf
fastElkies$'$} requires that $2,\ldots, 8\ell-5$ be units in
$\K$. This completes the proof of Theorem~\ref{thisog}.

\medskip
In the case of odd $\ell$, we can compute $g(x)$ instead of $D(x)$.
Accordingly, we modify the recurrence relations, and compute fewer
terms. Let $q_1,q_2,\dots$ denote the power sums of $g(x)$, so that $q_i
= p_i/2$. Then, the coefficients $h_i$ and the power sums $q_i$ are
related by the relation
\begin{equation}\label{fnpng}
h_i = (4i+2) q_{i+1} + (4i-2) A q_{i-1} + (4i-4) B q_{i-2}.
\end{equation}
To compute $g(x)$ using algorithm {\sf fastElkies}, it suffices to
compute $S(x) \bmod x^{\ell+1}$; then $T(x)$ and $U(x)$ are computed
modulo $x^{(\ell+1)/2}$. Similarly, in algorithm {\sf fastElkies$'$}, it
is enough to compute $S(x) \bmod x^{4\ell}$, and $T(x)$ and $U(x)$
modulo $x^{2\ell}$.


\section{Examples of isogeny computations}\label{sct:examples}

\subsection{Worked example} 
Since the case of $\ell$ odd is quite important in practice, we first
give an example of such a situation (see below for an
example with $\ell=6$). Let
$$E: y^2=x^3+ x+ 1 \quad \text{and} \quad \Es: y^2 = x^3+75
x+16$$ be defined over $\F_{101}$, with $\ell=11$ and $\sigma =
50$. Since $\ell$ is odd, we will compute the polynomial $g(x)$, which
has degree $5$. First, from the differential equation
$$(x^6 + x^4 + 1)S'(x)^2 = 1 + 75 S(x)^4 + 16
S(x)^6, \quad S(0)=0, \quad S'(0)=1$$ we infer the equalities
$$\begin{array}{rccl}
  &C&=& 1 + 100 \, x^4 + 100\, x^6 + x^8 + 2\, x^{10} + O(x^{11}), \\[3pt]
  &S &=& x+68\,x^{5}+66\,x^{7}+60\,x^{9}+84\,x^{11} + O(x^{12}), \\[3pt]
  \text{so that~} & T &=& 1+68\,x^2+66\,x^{3}+60\,x^{4}+84\,x^{5}+ O(x^{6}),\\[3pt]
  \text{and~}&
  T^2 &=& 1+35x\,^2+31x\,^3+98x\,^4+54x\,^5+ O(x\,^6), \\ [3pt]
  \text{~whence~}& {U} & = & 1+66 x\,^2+70 x\,^3+16 x\,^4+96 x\,^5+O(x\,^{6}).
\end{array}$$
We deduce
\begin{eqnarray*}
  \frac{N(x)}{D(x)}
  &=&
  x+\frac{66}{x}+\frac{70}{x^2}+\frac{16}{x^3}+\frac{96}{x^4} +O \left( \frac1{x^5} \right).
\end{eqnarray*}
At this stage, we know $h_1 =66, h_2=70, h_3=16,h_4=96$, as well 
as $q_1 = \sigma/2 = 25$.
Equation~\eqref{fnpng} then writes
$$q_{i+1} = \frac{h_i - (4i-2) q_{i-1} - (4i-4) q_{i-2}}{4i+2}, \quad
\text{for all} \; 1\leq i \leq 4$$ and gives $q_2=43, q_3=91,
q_4=86,q_5=63$. The main equation in \S\ref{ssec:exp} writes
\begin{eqnarray*}
  x^5 g \left ( \frac 1x \right ) &=& 
  \exp_{6}\!\left (
    -\left(25\,x + \frac{43}2 x^2 + \frac{91}3\, x^3 + \frac{86}4\, x^4+\frac{63}5\,x^5
    \right ) \right) \\
  &=&\exp_{6}\!\left (76\,x + 29\,x^2 + 37 \, x^2 + 29 \, x^4 +48 \,x^5 \right ),
\end{eqnarray*}
yielding $g(x) = x^5+76 x^4+89 x^3+24 x^2+97 x+5.$ For the sake
of completeness, we have:
$$N(x)=x^{11}+51 x^{10}+61 x^9+44 x^8+71 x^7+39 x^6+81 x^5+43 x^4+15
x^3+5 x^2+24 x+15.$$ Had we computed the solution $S(x)$ at precision
$O(x^{44})$, the expansion at infinity of $N(x)/D(x)$ would have been known
at precision $O(1/x^{21})$, and this would have sufficed to recover
both $N(x)$ and $D(x)$ by rational function reconstruction, without
the prior knowledge of $\sigma$.


\subsection{Further examples.}
As it turns out, all the theory developed for the prime case in the
SEA algorithm works also, {\it mutatis mutandis}, in the more general
case of a cyclic isogeny of non prime degree.  Consider the curve $$E
: y^2 = x^3+x+3$$ defined over $\GFq{1009}$. For $\ell=6$, we find
that the modular polynomial of degree $6$ (obtained as the resultant
of the modular polynomials of degrees $2$ and $3$) has three roots,
one of which is $\js = 248$. Using the formulas of \cite{BlSeSm99},
{\em that are still valid}, we find the isogenous curve
$$\Es: y^2 = x^3+830 x+82$$ and $\sigma = 739$,
from which we obtain
$$\frac{N(x)}{D(x)}= {\frac
{{x}^{6}+270\,{x}^{5}+325\,{x}^{4}+566\,{x}^{3}+382\,{x}^{2}+
555\,x+203}{{x}^{5}+270\,{x}^{4}+289\,{x}^{3}+659\,{x}^{2}+533\,x+399}
}.$$ 
The denominator  factors as
$$\left (x-66\right )\left (x-23\right )^{2}\left (x-818\right
)^{2}.$$ The value $x=66$ corresponds to one of the roots of $x^3+x+3$
and is therefore the abscissa of a point of $2$-torsion; $23$ is the
abscissa of a point of $3$-torsion; $818$ is the abscissa of a
primitive point of $6$-torsion.

As an aside, let us illustrate the case of a non cyclic isogeny. The
curve $E$ happens to have rational $2$-torsion; the subgroup $E[2]$ of
$2$-torsion is non cyclic, being isomorphic to $ \ZN{2} \times
\ZN{2}$. The denominator $D(x)$ appearing in the isogeny $I: E
\rightarrow \Es = E/E[2]$ is simply $D(x) = x^3+x+3$ and
Equation~\eqref{eqND} yields $N(x) =
{x}^{4}+1007\,{x}^{2}+985\,x+1$. From this, we can compute the
equation of $\Es$, namely, $y^2 = x^3 + 16 x + 192$.



\section{A survey of previous algorithms for isogenies}

In this section, we recall and give complexity results for other known
algorithms for computing isogenies. In what follows, we write the
Weierstrass functions $\wp$ and $\wps$ of our two curves $E$ and $\Es$
as
$$\wp(z) = \frac{1}{z^2} + \sum_{i\geq 1} c_i z^{2i} \quad \text{and}
\quad \wps(z) = \frac{1}{z^2} + \sum_{i\geq 1} \tilde{c}_i z^{2i}.$$
All algorithms below require the knowledge of the expansion of these
functions at least to precision $\ell$, so they only work under a
hypothesis of the type $p \gg \ell$ or $p=0$.

We can freely assume that these expansions are known. Indeed, by
Theorem~\ref{thwp}, given $A,B$ and $\As,\Bs$, we can precompute the
coefficients $c_i$ and $\tilde c_i$ up to (typically) $i=\ell-1$ using
$O(\sfM(\ell))$ operations in $\K$, provided that the characteristic
$p$ of $\K$ is either $0$ or $> \ell$. This turns out to be negligible
compared to the other costs involved in the following algorithms.


\subsection{First algorithms}
A brute force approach to compute $N(x)/D(x)$ is to use the equation
\begin{equation}\label{eq:wpwps}
\wps(z) = \wp \left(\frac{N(z)}{D(z)} \right)
\end{equation}
and the method of undetermined coefficients.  This reduces to
computing $\wp(z)^i \bmod {z^{4\ell - 2}}$ for $1\leq i \leq \ell$ and
solving a linear system with $2\ell-1$ unknowns. This direct method
requires that $2,\ldots, 4\ell$ be units in $\K$ and its complexity is
$O(\ell^\omega)$ operations in $\K$, where $2\leq \omega <3$ is the
exponent of matrix multiplication.

Another possible idea is to consider the rational functions
$N(x)/D(x)$ and $\hat{N}(x)/\hat{D}(x)$ respectively associated to $I$
and its dual $\hat I$, noticing that by definition,
$$
\frac{N}{D} \circ \frac{\hat{N}}{\hat{D}} =
\frac{\phi_\ell}{\psi_\ell^2},
$$ using the notation of Equation~\eqref{eq:divpol}.  However,
algorithms for directly decomposing
$\phi_\ell/\psi_\ell^2$~\cite{Zippel91,GuRe92,AlGuRe95} lead to an
expensive solution in our case, since they require factoring the
division polynomial $f_\ell$, of degree $\Theta(\ell^2)$. Indeed,
even using the best (sub-quadratic) algorithms for polynomial
factorization~\cite{KaSh98}, of exponent $1.815$, this would yield an
algorithm for computing isogenies of degree $\ell$ in complexity more
than cubic with respect to $\ell$, which is unacceptable.


\subsection{Stark's method}

To the best of our knowledge, the first subcubic method for finding
$N$ and $D$ is due to Stark \cite{Stark72} and amounts to expanding
$\wps$ as a continued fraction in $\wp$, using
Equation~\eqref{eq:wpwps}. The fraction $N/D$ is approximated by
$p_n/q_n$ and the algorithm stops when the degree of $q_n$ is
$\ell-1$, yielding $D$. In particular, it works for any degree
isogeny.  Since $\wp$ and $\wps$ are in $1/z^2+\K[[z^2]]$, it is
sufficient to work with series in $Z =z^2$.

\medskip

\begin{enumerate}
\item $T := \wps(Z) +O(Z^{\ell})$;
\item $n := 1$; 
\item $q_0:=1$;
\item $q_1:=0$;
\item {\bf while} $\mathrm{deg}(q_n) < \ell-1$ {\bf do}
\begin{enumerate}
\item []\hspace{-0.75cm}{\rm \{at this point, $T(Z) = t_{-r} Z^{-r} + \cdots+ t_0 + t_1 Z + \cdots+O(Z^{(\ell-\deg q_n-r)-1})$\}}
\item $n := n+1$;
\item $a_n := 0$; 
\item 
{\bf while} $r \geq 1$ {\bf do} 
\begin{itemize}
\item[] $a_n:=a_n +t_{-r} z^r;$
\item[] $T:=T - t_{-r} \wp^r=t_{-s}Z^{-s}+\dotsb;$
\item[]$r:=s$
\end{itemize}
\item $q_n := a_n q_{n-1} + q_{n-2}$;
\item $T := 1/T$;
\end{enumerate}
\item Return $D := q_n$.
\end{enumerate}

\medskip This algorithm (that we call {\sf Stark1972}) requires
$O(\ell)$ passes through Step (5); this bound is reached in general,
with $r=1$ at each step. The step that dominates the complexity is the
computation of reciprocals in Step (5.e), with
precision~$2\ell-1-2\deg q_n-2r$. The sum of these operations thus
costs~$O(\ell\sfM(\ell))$. The multiplications in Step (5.d) can be
done in time $O(\ell\sfM(\ell))$ as well (these multiplications could
be done faster if needed).  Since the largest degree of the
polynomials $a_n$ is bounded by $\ell-1$, computing all powers of
$\wp$ at Step (5.c) also fits within the $O(\ell\sfM(\ell))$ bound.
Finally, knowing $D(x)$, the numerator~$N(x)$ can be recovered in cost
$O(\sfM(\ell))$ using Equation~\eqref{eqND}.

In the case where $\ell$ is odd and we need $g(x)$, as in the context
of the SEA algorithm, we can compute it in $O(\sfM(\ell))$ operations
by computing~$\exp((\log D)/2)$.

To summarize, the total cost of algorithm {\sf Stark1972} is in
$O(\ell\sfM(\ell))$. Remark that compared to the methods presented
below, algorithm {\sf Stark1972} does not require the knowledge of
$\sigma$. Remark also that, even though $r$ will
be $1$ in general, the computation of the powers $\wp^r$ in Step (5.c)
could be amortized in the context of the SEA algorithm.


\medskip
\subsection{Elkies' 1992 method}~\label{elkies92} We reproduce the
method given in~\cite{Elkies92}, that we call {\sf Elkies1992} (see
also e.g.,~\cite{ChCoRo91,Morain95a}).  We suppose that $\ell$ is odd,
so that $D(x)=g(x)^2$, though minor modifications below would lead to
the general solution.

Differentiating twice Equation~\eqref{eqdifff} yields
$$\frac{d^4 \wp(z)}{d z^4} = 120 \wp^3 + 72 A \wp + 48 B.$$
More generally, we obtain equalities of the form
$$\frac{d^{2k} \wp(z)}{d z^{2k}} = \mu_{k,k+1} \wp^{k+1} + \cdots +
  \mu_{k,0},$$ for some constants $\mu_{k,j}$ that satisfy the recurrence relation
\begin{equation*}\label{eq:mu}
  \mu_{k+1,j} =   (2j-2) (2j-1) \mu_{k,j-1} +  (2j+1) (2j+2) A \mu_{k,j+1}
  + (2j+2)(2j+4) B \mu_{k,j+2},
\end{equation*}
with $\mu_{k,k+1} = (2 k+1)!$. Using this recurrence relation, the
coefficients $\mu_{k,j}$, for $k \leq d-1$ and $j \leq k+1$, can be
computed in $O(\ell^2)$ operations in $\K$.

Elkies then showed how to use these coefficients to recover the power
sums $q_2,\dots,q_d$ of $g$, through the following equalities, 
holding for $k \geq 1$:
$$(2 k)! (\tilde{c}_k - c_k) = 2 (\mu_{k,0} q_0 + \cdots + \mu_{k,k+1}
q_{k+1}).$$ Using these equalities, assuming that $q_1=\sigma/2$ and the
coefficients $c_k$, $\tilde c_k$ and $\mu_{k,j}$ are known, we can
recover $q_2,\dots,q_d$ by solving a triangular system, in complexity
$O(\ell^2)$. We can then recover $g$ using either a quadratic
algorithm, or the faster algorithm of \S\ref{ssec:exp}.

There remains here the question whether the triangular system giving
$q_2,\dots,q_d$ can be solved in quasi-linear time. To do so, one
should exploit the structure of the triangular system, so as to avoid
 computing the $\Theta(\ell^2)$~constants $\mu_{k,j}$ explicitly.


\subsection{Atkin's method}\label{ssec:Atkin}
In~\cite{Atkin92b}, Atkin gave a formula enabling the computation of
$D(x)$ (see also~\cite[Formula 6.13]{Mueller95} and~\cite{Schoof95})
in the case where $\ell$ is odd. We extend it, so as to cover the case
of arbitrary $\ell$, this time recovering $D(x)$. The equation we use
is
\begin{equation}\label{gwp} 
  D(\wp(z)) = z^{2 - 2\ell} \exp(F(z)),  
\end{equation}
where
$$F(z) = - \sigma z^2 +2 \left(\sum_{k=1}^{\infty} (\ell c_k -
  \tilde{c}_k) \frac{z^{2k+2}}{(2k+1) (2k+2)}\right).$$ Since $\ell$
  and the coefficients $c_k,\tilde{c}_k$ are all assumed to be known,
  one can deduce the series $F(z) \bmod z^\ell$, provided that $\sigma$
  is given. A direct method to determine $D(x)$ is then to compute the
  exponential of $F(z)$, and to recover the coefficients of $D(x)$ one at a
  time, as shown in the following algorithm, called {\sf
  Atkin1992}. As before, we use series in $Z = z^2$.

\begin{enumerate}
\item Compute the series $P_i(Z) = \wp(Z)^i$ at order $\ell$, for $1\leq i\leq \ell-1$;
\item Compute 
  $G(Z) = \exp_{\ell}(F(Z))$; 
\item $T:=G$;
\item $D:=0$;
\item {\bf for} $i:=\ell-1$ {\bf downto} $0$ {\bf do}
\begin{enumerate}
\item []\hspace{-0.75cm} \{at this point, $T = t Z^{-i} + \cdots$ \}
\item $D := D + t z^i$;
\item $T := T - t P_i$.
\end{enumerate}
\end{enumerate}
Step (1) uses $O(\ell \sfM(\ell))$ operations; the cost of Step (2) is
negligible, using either classical or fast exponentiation. Then, each
pass through Step (5) costs $O(\ell)$ more operations, for a total of
$O(\ell^2)$. Thus, the total cost of this algorithm is in $O(\ell
\sfM(\ell))$.  If this algorithm is used in the context of the SEA
algorithm, Step (1) can be amortized, since it depends on $E$
only. Therefore, all the powers of $\wp$ should be computed for the
maximal value of $\ell$ to be used, and stored. Hence, the cost of
this algorithm would be dominated by that of Step~(5), yielding a
method of complexity $O(\ell^2)$.

A better algorithm for computing $D(x)$, avoiding the computation of all
powers of $\wp(z)$, is based on the remark that Equation~\eqref{gwp}
rewrites
\begin{equation}\label{eq:expF}
  D\left ( \frac 1x \right ) = 
  \mathcal{I}^{2-2\ell} \left( (\exp \circ F) \circ \mathcal{I}
  \right),
\end{equation}
with $\mathcal{I}(x) = \wp^{-1}(1/x)$, where $\wp^{-1}$ is the
functional inverse of $\wp$. The expansion of $ \mathcal{I}(x)$ at order
$\Theta(\ell)$ can be computed in $O(\ell)$ operations using the
differential equation
\begin{equation}\label{eq:I}
  \mathcal{I}'(x)^2=\frac{1}{4x(1+Ax^2+Bx^3)} 
  \quad \text{or}\quad
  \mathcal{I}'(x)=\frac{1}{2\sqrt{x}} \frac 1{\sqrt{1+Ax^2+Bx^3}}.
\end{equation}
A linear differential equation follows:
\[\frac{\mathcal{I'}(x)}{\mathcal{I}(x)}=-\frac{1+3Ax^2+4Bx^3}{2x(1+Ax^2+Bx^3)}.\]
{}From there follows a linear differential equation
for~$\mathcal{J}(x)=x^{-1/2}\mathcal{I}(x)=\sum_{i\ge0}{a_ix^i}$.
Extracting coefficients in this equation then gives a linear
recurrence
\begin{equation}\label{eq:A}
a_{i+1}=-\frac{2i-1}{2(i+1)(2i+3)}\left((2i-3)Ba_{i-2}+2Aia_{i-1}\right)
\quad  \text{for $i \geq 2$,}
\end{equation}
with initial conditions $\; a_0=1, a_1=0, a_2=-\frac{A}{10}.$
This yields the following algorithm, called {\sf AtkinModComp}:
\begin{enumerate}
\item Compute $G(Z) = \exp_{\ell}(F(Z))$;
\item Compute $\mathcal{I}(x)$ using Equation~\eqref{eq:A};
\item Compute $G(\mathcal{I})$ by modular composition (which is
  possible since $G$ is in $\K[[Z]]=\K[[x^2]]$);
\item Deduce $D$ using Equation~\eqref{eq:expF}.
\end{enumerate}
The cost of the algorithm is dominated by the composition of the
series $G=\exp \circ F$ and $\mathcal{I}$. From~\S\ref{ssec:other},
this can be done in $O({\sf M}(\ell) \sqrt{\ell} + \ell^{\frac
  {\omega+1}2} )$ or $O({\sf M}(\ell) \, \sqrt{\ell \log \ell} )$
operations in $\K$.

To do even better, it is fruitful to reconsider the series
$G(\mathcal{I})= (\exp \circ\, F) \circ \mathcal{I}$ used above, but
rewriting it as $\exp \circ\, ( F \circ \mathcal{I})$ instead; this
change of point of view reveals close connections with our {\sf
  fastElkies} algorithm. More precisely, Atkin's Equation~\eqref{gwp}
can be rewritten as
\begin{equation*}\label{eq:Atkin2}
D({\wp}(x)) = \exp\left(-\sigma x^2 +2 \iint \ell \wp(x) -
\wps(x)\right).
\end{equation*}
We can then obtain $D(1/x)$ as the following exponential:
\begin{eqnarray} D\left( \frac 1x \right ) &=& \exp \left (-\sigma
    {\mathcal{I}}^2 +2 \int {\mathcal{I}}' \int {\mathcal{I}}' \left
      (\frac \ell x -(\wps \circ \mathcal{I})(
      x ) \right )\right )\\
  &=& \exp \left (-\sigma {\mathcal{I}}^2 + 2\int {\mathcal{I}}' \int
    {\mathcal{I}}' \left (\frac \ell x -\frac {N(1/x)}{D(1/x)} \right )\right )\label{eq:2} \\
  &=& \exp \left (-\sigma {\mathcal{I}}^2 + 2\int{\mathcal{I}}' \int
    {\mathcal{I}}' \left (\frac \ell x - \frac 1{S( \sqrt{x} )^2}
    \right ) \right ).
   \end{eqnarray}
   Then, working out the details, the sequence of operations necessary
   to evaluate this exponential turns out to be the same as the one
   used in our algorithm {\sf fastElkies} of \S\ref{fastElkies}.  This
   does not come as a surprise: the relation~\eqref{fnpn} used in our
   algorithm follows from formula~\eqref{eqND}, which can be
   rewritten as
\begin{equation*}
  \frac{N(x)}{D(x)}=\ell x-\sigma- 2 \sqrt{x^3+A x+B} \left(\sqrt{x^3+A
      x+B}\, \frac{D'(x)}{D(x)}\right)'.
\end{equation*}
Then, Equation~\eqref{eq:2} is nothing but an integral reformulation
of this last equation, taking into account the fact that $\mathcal{I}$
satisfies the differential equation~\eqref{eq:I}.


\medskip
\subsection{Summary}
In Table~\ref{table} we gather the various algorithms discussed in
this article, and compare these algorithms from two points of view:
their complexity (expressed in number of operations in the base field
$\K$) and their need for $\sigma$ as input.

\begin{table} [ht]
$$\begin{array}{|c||c|c|}\hline
  \text{algorithm} & \text{complexity} & \text{need of}\; \sigma \\ 
   \hline
  \text{linear algebra}      & O(\ell^\omega)    & \text{no}  \\ 
  \textsf{Stark1972}       & O(\ell \sfM (\ell))   & \text{no} \\ 
  \textsf{Atkin1992}       & O(\ell \sfM (\ell))  & \text{yes} \\ 
  \textsf{AtkinModComp}  & O({\sf M}(\ell) \sqrt{\ell} + \ell^{\frac
    {\omega+1}2} ) \text{~or~} O({\sf M}(\ell) \, \sqrt{\ell \log \ell} )  & \text{yes} \\
  \textsf{Elkies1992}    & O(\ell^2) & \text{yes}  \\
  \textsf{Elkies1998}    & O(\ell^2) & \text{yes} \\
  \textsf{fastElkies}  & O(\sfM (\ell)) &  \text{yes} \\
  \textsf{fastElkies$'$} & O(\sfM (\ell) \log \ell) & \text{no} \\
  \hline
\end{array}$$
\caption{\label{table}Comparison of the algorithms}
\end{table}


\section{Implementation and benchmarks}\label{section:implementation}
\begin{figure}[htb]
\centerline{  \includegraphics[scale=0.3,angle=270]{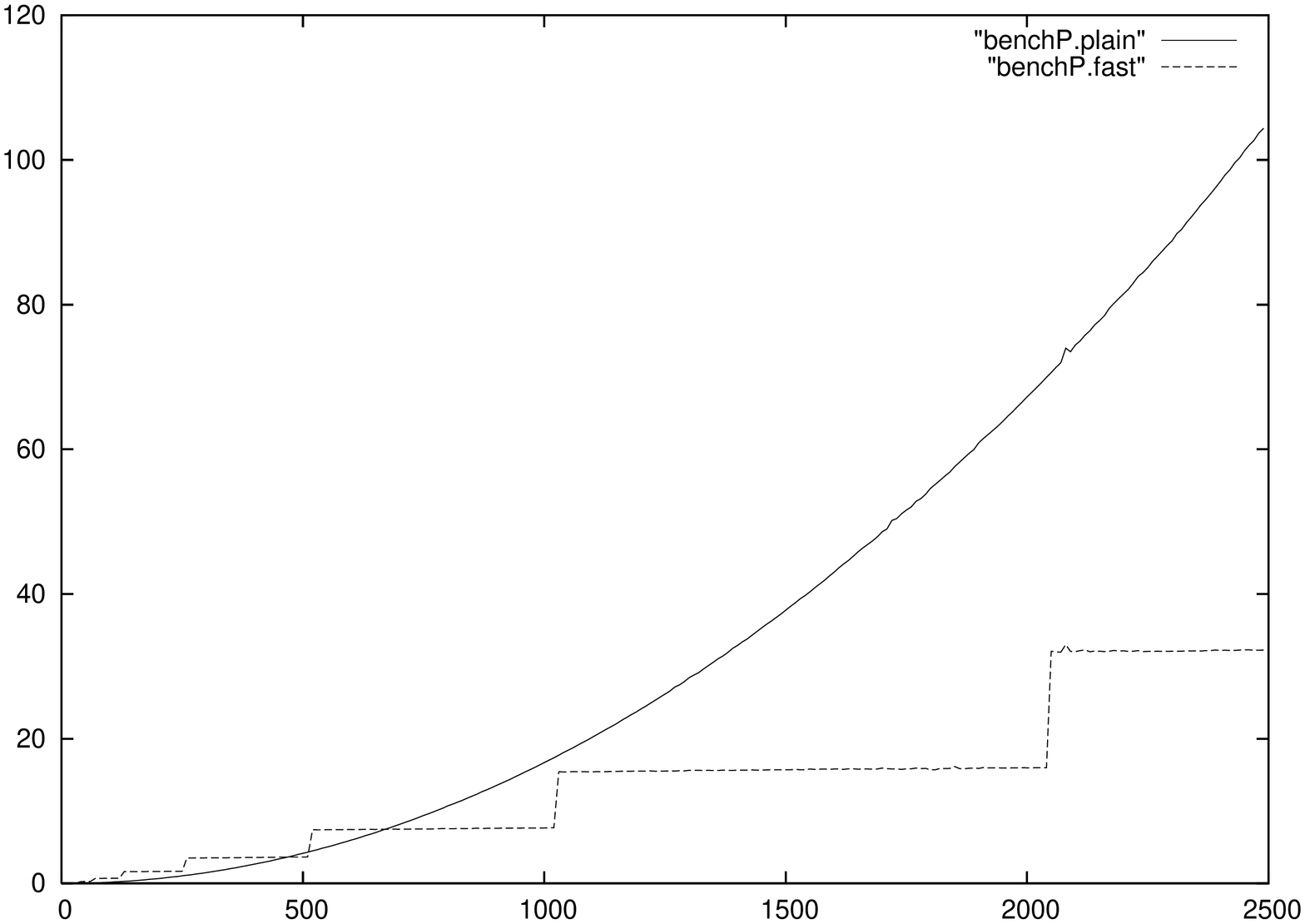}}
  \caption{Timings for computing $\wp$ on $E: y^2 = x^3+4589 x+91128$
    over $\GFq{10^{2004}+4683}$\label{tablewp}}
\end{figure}

We implemented our algorithms using the NTL C++
library~\cite{Shoup95,NTL} and ran the program on an AMD 64 Processor
3400+ (2.4GHz). We begin with timings for computing the expansion of
$\wp$, obtained over the finite field $\GFq{10^{2004}+4683}$; they are
given in Figure~\ref{tablewp}.

The shape of both curves indicates that the
theoretical complexities -- quadratic vs. nearly linear -- are well
respected in our implementation (note that the abrupt jumps at powers
of~2 reflect the performance of NTL's FFT implementation of polynomial
arithmetic). Moreover, the threshold beyond which our algorithm
becomes useful over the quadratic one is reasonably small, making it
interesting in practice very early.

We now turn our attention to the pure isogeny part, concentrating on
the case where $\ell$ is prime, in the context of the SEA algorithm.
Hence, in this case, it suffices to compute the polynomial $g(x)$ such
that $D(x)=g(x)^2$. All algorithms can be adapted to make advantage of
this simplification, as exemplified in~\S\ref{fastElkies} for our 
algorithms {\sf fastElkies} and {\sf fastElkies$'$}.

\begin{figure}[!!h]
\centerline{\includegraphics[scale=0.3,angle=270]{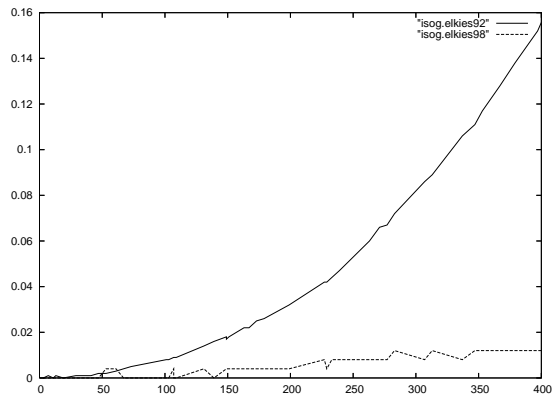}}
\caption{{\sf Elkies1992} vs. {\sf Elkies1998}. \label{figone}}
\end{figure}

\begin{figure}[!!h]
\centerline{\includegraphics[scale=0.3,angle=270]{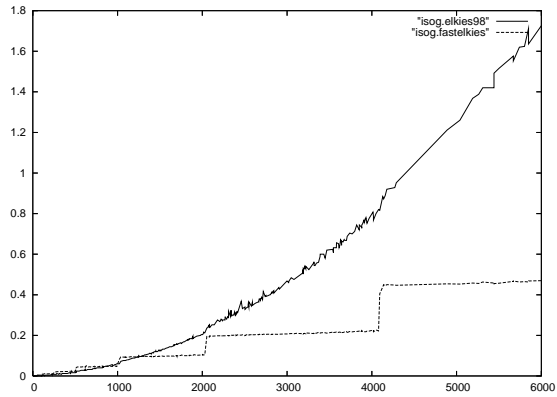}}
\caption{{\sf Elkies1998} vs. {\sf fastElkies}. \label{figtwo}}
\end{figure}

The first series of timings concerns the computation of isogenies over
a small field, $\K=\GFq{10^{19}+51}$, for the curve $E: y^2 = x^3+4589
x+91128$. We compare in Figure~\ref{figone} the performances of the
algorithms {\sf Elkies1992} from~\S\ref{elkies92} and {\sf Elkies1998}
from~\S\ref{elkies98} for isogenies of moderate degree $\ell \leq
400$.  Figure~\ref{figtwo} compares the timings obtained with the
algorithm {\sf Elkies1998} and our fast version \textsf{fastElkies}
from~\S\ref{fastElkies}, for isogenies of degree up to $6000$.

\begin{figure}[!!h]
\begin{center}
\centerline{\includegraphics[scale=0.3,angle=270]{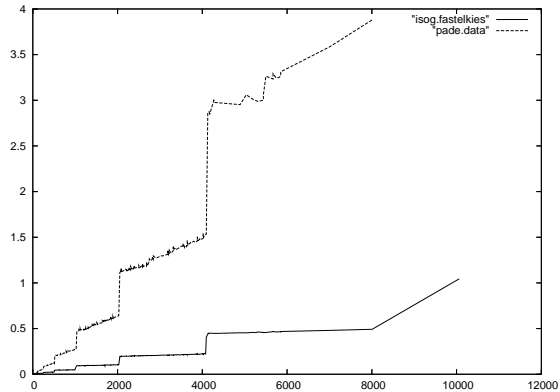}}
\caption{{\sf FastElkies} vs. {\sf FastElkies$'$} \label{figthree}}
\end{center}
\end{figure}

Next, we compare in Figure~\ref{figthree} the timings obtained by the
$O(\sfM(\ell))$ algorithm {\sf fastElkies}, that requires the
knowledge of $\sigma$, to those obtained by its $O(\sfM(\ell)\log
\ell)$ counterpart {\sf fastElkies$'$}, that does not require this
information.

In all figures, the degrees $\ell$ of the isogenies are represented on
the horizontal axis and the timings are given (in seconds) on the
vertical axis. Again, the shape of both curves in Figure~\ref{figtwo}
shows that the theoretical complexities are well respected in our
implementation. The curves in Figure~\ref{figthree} show that the
theoretical ratio of $\log \ell$ between algorithms {\sf fastElkies}
and {\sf fastElkies$'$} has a consequent practical impact.


Next, in Tables~\ref{tab:wp} to~\ref{tab:Stark}, we give detailed
timings on computing $\ell$-isogenies for the curve $$E:
y^2 = x^3+A x+B$$
where
$$A = \lfloor 10^{1990} \pi\rfloor = 31415926\ldots 58133904,$$
$$B = \lfloor 10^{1990} e\rfloor = 27182818\ldots 94787610,$$
for a few values of $\ell$, over the larger
finite field $\GFq{10^{2004}+4683}$, and using various methods:
algorithms {\sf Elkies1992}, {\sf Elkies1998} and  our fast variant
{\sf fastElkies}, Stark's algorithm {\sf Stark1972} and the two versions
{\sf Atkin1992} and {\sf AtkinModComp} of Atkin's algorithm.

Tables~\ref{tab:wp} and~\ref{tab:g} give timings for basic subroutines
shared by some or all of the algorithms discussed. Table~\ref{tab:wp}
gives the timings necessary to compute the expansions of $\wp$ and
$\wps$, using either the classical algorithm or our faster variant:
this is used in all algorithms, except our {\sf fastElkies} algorithm.
Table~\ref{tab:g} gives timings for recovering $g$ from its power
sums, first using the classical quadratic algorithm, and then using
fast exponentiation as described in \S\ref{ssec:exp}. This is used in
algorithms {\sf Elkies1992}, and {\sf Elkies1998} and its variants.

\begin{table}[!!h]
\begin{minipage}[b]{0.47\textwidth}
$$
\begin{array}{|c||r|r|r|}\hline
  & \multicolumn{3}{c|}{\text{Computing $\wp$ and $\wps$}} \\
  \ell & \text{order} & \text{quadratic} & \text{fast} \\  \hline
  1013 &  511 &  8.6 & 7.0 \\
  2039 & 1024 & 34.6 & 29.9 \\
  3019 & 1514 & 75.7 & 30.3 \\
  4001 & 2005 & 132.7 & 31\\
  5021 & 2515 & 209.3 & 64.4\\
  \hline
\end{array}
$$
\caption{Computing $\wp$ and $\wps$\label{tab:wp}}
\end{minipage}
\begin{minipage}[b]{0.49\textwidth}
$$
\begin{array}{|c||r|r|}\hline
  & \multicolumn{2}{c|}{\text{Recovering $g$}} \\
  \ell &  \text{quadratic} & \text{fast} \\  \hline
  1013 &   4.2 &  1.1 \\
  2039 &  17.4 &  2.5 \\
  3019 &  38.2 &  5.1 \\
  4001 &  66.9 &  5.5 \\
  5021 & 106.2 & 11.2 \\
  \hline
\end{array}
$$
\caption{Recovering $g$ from its power sums\label{tab:g}}
\end{minipage}
\end{table}

Tables~\ref{tab:92} and~\ref{tab:98} give the timings for algorithms
{\sf Elkies1992} on the one hand and {\sf Elkies1998} and our
variation {\sf fastElkies} on the other hand.  In Table~\ref{tab:92},
the columns $\mu$ and $p_i$ give the time used to compute the
coefficients $\mu_{i,j}$ and the power sums $p_i$.  In
Table~\ref{tab:98}, the column $h_i$ indicates the time used to
compute the coefficients $h_i$ of the rational function $N/D$, first
using the original quadratic algorithm {\sf Elkies1998}, then using
our faster variant {\sf fastElkies}. The next column gives the time
used to compute the power sums~$p_i$ from the $h_i$ using
recurrence~\eqref{fnpng}.

\begin{table}[!!!h]
$$
\begin{array}{|c||c|r|r|c|}\hline
  & \multicolumn{4}{c|}{\textsf{Elkies1992}} \\
  \ell  & \multicolumn{1}{c|}{\wp,\wps} & \multicolumn{1}{c|}{\mu} &  \multicolumn{1}{c|}{p_i} &  g \\ \hline
  1013      &                           & 10.4   & 4.4     & \\
  2039      & \text{See}                & 49.1 & 17.9   & \text{See} \\
  3019      & \text{Table~\ref{tab:wp}} & 130.6 & 38.9   & \text{Table~\ref{tab:g}}\\
  4001      &                           & 263 & 68.4  & \\
  5021      &                           & 496.5 & 106.6 & \\
  \hline
\end{array}
$$
\caption{Algorithm {\sf Elkies1992}\label{tab:92}}
\end{table}

\begin{table}[!!!h]
$$
\begin{array}{|c||r|r|r|c|}\hline
  & \multicolumn{4}{c|}{\textsf{Elkies1998} \text{~and~} \textsf{fastElkies}}  \\
  \ell & \multicolumn{2}{c|}{h_i} & \multicolumn{1}{c|}{p_i} & g  \\
  & \text{quadratic}  & \text{fast} &       &  \\ \hline
  1013 & 4.4               &  4.5        &  0.05 &  \\
  2039 & 17.3              &  9.6        &  0.1 & \text{See} \\
  3019 & 38.0              & 19.5        &  0.16 & \text{Table~\ref{tab:g}} \\
  4001 & 67.2              & 20.0        &  0.21 &  \\
  5021 & 105.0             & 40.7        &  0.27 &  \\
  \hline
\end{array}
$$
\caption{Algorithms {\sf Elkies1998} and \textsf{fastElkies}\label{tab:98}}
\end{table}

Tables~\ref{tab:Atkin} and~\ref{tab:modcomp} give timings for our
implementation of Atkin's original algorithm {\sf Atkin1992}, as well
as the faster version {\sf AtkinModComp} using modular composition
mentioned in \S\ref{ssec:Atkin}.  
In Table~\ref{tab:Atkin}, the column
``exponential'' compares the computation of $\exp(F)$ using the naive
exponentiation algorithm to the computation using the faster algorithm
presented in \S\ref{ssec:exp}; the column $\wp^k$ gives the time for
computing all the series $\wp(z)^k$ and the column $g$ that for
recovering the coefficients of $g$ from its power sums.
Table~\ref{tab:modcomp} gives timings obtained using the two modular
composition algorithms mentioned in~\S\ref{ssec:other}, called here
\textsf{ModComp1} and \textsf{ModComp2}; the previous columns give the
time for computing $\exp(F)$ and that for computing the requested
power of~$\mathcal{I}$; the last column gives the time to perform the
final multiplication.

\begin{table}[!!!h]
$$
\begin{array}{|c||c|r|r|r|r|}\hline
  &  \multicolumn{5}{c|}{\text{Algorithm {\sf Aktin1992}}} \\
  \ell & \wp,\wps & \multicolumn{2}{c|}{\text{exponential}}   &  \multicolumn{1}{c|}{\wp^k} &   \multicolumn{1}{c|}{g}  \\
  &                           & \text{naive} & \text{fast}  &  &   \\  \hline
  1013 &                           &  88.4 & 1.2   & 72.3 &  4.4 \\
  2039 & \text{See}                &  370.1 & 4.9   & 304.9&  17.7 \\
  3019 & \text{Table~\ref{tab:wp}} &  955.9 & 5.1   & 755.8&  38.9 \\
  4001 &                           & 1503 & 5.2   & 1218.9&  67.6  \\
  5021 &                           & 3180 & 10.8 & 2506.4 & 108.7  \\
  \hline
\end{array}
$$
\caption{Atkin's original algorithm, variations for $\exp(F)$ \label{tab:Atkin}}
$$
\begin{array}{|c||c|r|r|r|r|r|}\hline
  & \multicolumn{6}{c|}{\text{Algorithm {\sf AtkinModComp}}} \\
  \ell & \wp,\wps & \exp(F) & \mathcal{I}^{1-\ell}  & \multicolumn{2}{c|}{\text{modular composition}}&g \\
  & &&         & \textsf{ModComp1} & \textsf{ModComp2} & \\ \hline
  1013  &                           & 1.2 & 2.7 & 14.3  & 35.6 & 0.2 \\
  2039  & \text{See}                & 2.5 & 6.6 & 45.8  & 111.9 & 0.4\\
  3019  & \text{Table~\ref{tab:wp}} & 5.1 & 10.4  & 95.3 & 241 & 0.7 \\
  4001  &                           & 5.2 & 11.6  & 143.2 & 338 & 0.9 \\
  5021  &                           & 10.9& 20.9  & 240 & 642 & 1.4\\
  \hline
\end{array}$$
\caption{Atkin's algorithm with modular composition\label{tab:modcomp}}
\end{table}

Asymptotically, algorithm {\sf ModComp2} is
faster than algorithm {\sf ModComp1}, so that the timings in
Table~\ref{tab:modcomp} might come as a surprise. The explanation is
that, for the problem sizes we are interested in, the predominant step
of algorithm {\sf ModComp1} is the one based on polynomial operations,
while the step based on linear algebra operations takes only about
10\% of the whole computing time.  Thus, the practical complexity of
this algorithm in the considered range ($1000 < \ell < 6000$) is
proportional to ${\sf M} (\ell) \sqrt{\ell}$, while that of algorithm
{\sf ModComp2} is proportional to ${\sf M} (\ell) \sqrt{\ell \log
\ell}$. Moreover, the proportionality constant is smaller in the
built-in NTL function performing {\sf ModComp1} than in our
implementation of {\sf ModComp2}.

Notice that in all the columns labelled ``fast'' in
Tables~\ref{tab:wp}--\ref{tab:modcomp}, the timings reflect the already
mentioned (piecewisely almost constant) behaviour of the FFT:
polynomial multiplication in the degree range 1024--2047 is roughly twice as fast
as in the range 2047--4095 and roughly four times as fast as in the range
4096--8191. 

Finally, Table~\ref{tab:Stark} gives timings for Stark's algorithm
{\sf Stark1972}; apart from the common computation of $\wp$ and
$\wps$, we distinguish the time necessary to compute all inverses
(the quadratic algorithm when available, followed by that using fast
 inversion)
and that for deducing the polynomials $q_n$.
\begin{table}[!!h]
$$\begin{array}{|r||c|r|r|r|}\hline
\ell & \wp,\wps          & \multicolumn{2}{c|}{\text{Inverses}} & q_n  \\ 
     &                   & \text{quadratic} & \text{fast}  &       \\ \hline
1013 &                           & 23542& 1222.7 & 28.0   \\
2039 & \text{See}                & >100000&5113.4    & 116.9  \\
3019 & \text{Table~\ref{tab:wp}} & &12182     & 258  \\
4001 &                           & &20388    & 418.6  \\
5021 &                           & &38910    & 663.1  \\ \hline
\end{array}$$
\caption{Stark's algorithm {\sf Stark1972}\label{tab:Stark}}
\end{table}





\section*{Conclusion}
The complexity analyses of the algorithms we have surveyed shows that
for the case of a large prime characteristic and for a reasonably
large degree~$\ell$ of the isogeny, our new $O(\sfM(\ell))$ algorithm
improves over previously known techniques.

The current implementation of our algorithm can be further optimized
to make it the algorithm of choice for smaller values of the
degree. Indeed, it is known that algorithms based on Newton iteration
present certain redundancies (coefficients that can be predicted in
advance, repeated multiplicands). Removing these redundancies is
feasible (see~\cite{Bernstein,HaQuZi04}), allowing one to achieve
constant-factor speed-ups.  For the moment, our implementation relies
only partially on these techniques; we believe that further
programming effort would bring practical improvements by
non-negligible constant factors.

Another direction for future work is to adapt our ideas to the case of
a small characteristic. In this respect, modifying the last phase of
the algorithm of Joux and Lercier \cite{JoLe06} seems a promising
search path.

\medskip\noindent {\bf Acknowledgments.} We thank Pierrick Gaudry for
his remarks during the elaboration of the ideas contained in this
work.

\bibliographystyle{plain}
\bibliography{BoMoSaSc06}

\end{document}